%% file: document.tex
\documentclass[letterpaper,11pt]{llncs}
\pdfoutput=1
\usepackage{algorithm,graphicx}
\usepackage{algpseudocode}
\usepackage{fixltx2e} 
\usepackage{amsmath}
\usepackage{soul}
\usepackage{colortbl}
\usepackage{sidecap}
\usepackage{todonotes}
\usepackage{url}
\usepackage{soul,multirow}
\usepackage{sidecap}
\usepackage{fullpage}

\def\sequel{\mbox{\rm {\sc misSEQuel}}}
\def\suffix{\mbox{\rm {\sf suffix}}}
\def\prefix{\mbox{\rm {\sf prefix}}}

\begin{document}

\title{Misassembly Detection using Paired-End Sequence Reads \\ and Optical Mapping Data}
\author{
Martin D. Muggli\inst{1}
\and
Simon J. Puglisi\inst{2}
\and
Roy Ronen\inst{3}
\and
Christina Boucher\inst{1}
}

\institute{
Department of Computer Science,\\
Colorado State University, Fort Collins, CO.\\
\email{\{muggli,cboucher\}@cs.colostate.edu}\\[1ex]
\and
Department of Computer Science,\\
University of Helsinki, Finland\\
\email{puglisi@cs.helsinki.fi}\\[1ex]
\and
Department of Computer Science and Engineering,\\
University of California, San Diego\\
\email{rronen@cs.ucsd.edu}\\[1ex]
}

\maketitle

\input{abstract}
\input{introduction}

\input{methods}
\input{results}

\input{conclusion}

\section*{Acknowledgements} 
The authors would like to thank Pavel Pevzner from the University of California, San Diego and Anton Korobeynikov from Saint Petersburg State University for many insightful discussions.  In addition, we would like to thank Alexey Gurevich from Saint Petersburg State University for clarifications and support with Quast.  

MM and CB were funded by the Colorado Clinical and Translational Sciences Institute which is funded by National Institutes of Health (NIH-NCATS,UL1TR001082, TL1TR001081, KL2TR001080).  SJP was supported by the Helsinki Institute of Information Technology (HIIT) and by Academy of Finland through grants 258308 and 250345 (CoECGR).  All authors greatly appreciate the funding provided for this project.  

\bibliographystyle{unsrt}
\bibliography{document}
\newpage
\input{appendix}
\end{document}

%% file: abstract.tex
\begin{abstract} 
A crucial problem in genome assembly is the discovery and correction of misassembly errors in draft genomes.  
We develop a method that will enhance the quality of draft genomes by identifying and removing misassembly errors using paired short read sequence data and optical mapping data.  
We apply our method to various assemblies of the loblolly pine and {\em Francisella tularensis} genomes.  
Our results demonstrate that we detect more than 54\% of extensively misassembled contigs and more than 60\% of locally misassembed contigs in an assembly of {\em Francisella tularensis}, and between 31\% and 100\% of extensively misassembled contigs and between 57\% and 73 \% of locally misassembed contigs in the assemblies of loblolly pine.  
$\sequel$ can be downloaded at \url{http://www.cs.colostate.edu/seq/}.
\end{abstract}

%% file: introduction.tex
\section{Introduction} \label{sec:intro}
Comparing genetic variation between and within a species is a fundamental activity in biological research. For example,  there is currently a major effort to sequence entire genomes of agriculturally important plant species to identify parts of the genome variable in a given breeding program and, ultimately, create superior plant varieties. Robust genome assembly methods are imperative to these large sequencing initiatives and other scientific projects~\cite{haussler2008genome,robinson2011creating,1001_arabidopsis,hmp} because scientific analyses frequently use those genomes to determine genetic variation and associated biological traits. 

At present, the majority of assembly programs are based on the Eulerian assembly paradigm~\cite{IW95,PTW}, where a de Bruijn graph is constructed with a vertex $v$ for every $(k - 1)$-mer present in a set of reads, and an edge $(v, v')$ for every observed $k$-mer in the reads with $(k - 1)$-mer prefix $v$ and $(k - 1)$-mer suffix $v'$. A contig corresponds to a non-branching path through this graph. We refer the reader to Compeau et al.~\cite{compeau} for a more thorough explanation of de Bruijn graphs and their use in assembly.  The assemblers Euler-SR \cite{Chaisson:2008}, Velvet \cite{Zerbino:2008}, SOAPdenovo \cite{soap}, ABySS \cite{Simpson:2009} and ALLPATHS \cite{Butler:2008} all use this paradigm and follow the same general outline: extract $k$-mers from the reads, construct a de Bruijn graph from the set $k$-mers, simplify the graph, and construct contigs.  

One crucial problem that persists in Eulerian assembly (and genome assembly, in general) is the discovery and correction of misassembly errors in draft genomes.  
We define a {\em misassembly error} as an assembled region that contains a significantly large insertion, deletion, inversion, or rearrangment that is the result of decisions made by the assembly program.  Identification of misassembly errors is important because true biological variations manifest in similar ways and thus, these errors can be easily misconstrued as true genetic variation~\cite{salzberg}. This can mislead a range of genomic analyses.  
We note that the exact definition of a misassembly error can vary, and adopt the standard definition used by Quast~\cite{quast} and other tools.  See section \ref{subsec:data} for this exact definition.  
Once the existence and 
location of a misassembly 
is identified, 
it
can be removed by segmenting the contig at that location.


We present a computational method for identifying misassembly errors using a combination of short reads and optical mapping data.   Optical mapping is a system developed in 1993~\cite{schwartz93} that can construct ordered, genome-wide, high-resolution restriction maps.  The system works as follows \cite{ORMenc,microfluidic}: an ensemble of DNA molecules adhered to a charged glass plate are elongated by fluid flow.   An enzyme is then used to cleave them into fragments at loci where the corresponding recognition sequence occurs. Next, the fragments are highlighted with fluorescent dye and imaged under a microscope. Finally, these images are analyzed to estimate the fragment sizes, producing a molecular map. Since the fragments stay relatively stationary during the aforementioned process, the images capture their relative order and size~\cite{Neely11}.   Multiple copies of the genome undergo this process, and a consensus map is formed that consists of an ordered sequence of fragment sizes, each indicating the approximate number of bases between occurrences of the recognition sequence in the genome \cite{Anantharaman01}.  

Although optical mapping data has been used for discerning structural variation in the human genome  \cite{teague}, and for scaffolding and validating contigs for several large sequencing projects --- including those for various prokaryote species \cite{reslewic,zhou,zhou2}, rice~\cite{RICE}, maize \cite{Zhou09}, mouse \cite{church}, goat~\cite{GOAT}, parrot~\cite{gigadb}, and {\em Amborella trichopoda} \cite{amborella} --- there exists no publicly available tools for using this data for misassembly correction using short read and optical mapping data.

Our tool, which we call $\sequel$, predicts which contigs are misassembled and the approximate locations of the errors in the contigs.  It takes as input the paired-end sequence read data, contigs, an ensemble of optical maps, and the restriction enzymes used to construct the optical maps.
$\sequel$ first uses the paired-end read data to divide the contigs into two sets: those that are predicted to be correctly assembled and those that are not.  
Then the set of  contigs that are candidates for containing misassembly errors are further divided into misassembled contigs and correctly assembled contigs using optical mapping data.
Fundamental to the first step is the concept of a {\em red-black positional de Bruijn graph}, which encapsulates recurring artifacts in the alignment of the sequence read data to the contigs and their position in the contig. 
The red vertices in this graph indicate if a contig is likely to be misassembled and also flag the location where the misassembly error occurs. These locations are called {\em  misassembly breakpoints}.

In the second stage of $\sequel$ where optical mapping data is used, the contigs conjectured to be misassembled are {\em in silico} digested with the set of input restriction enzymes and aligned to the optical map using Twin~\cite{wabi2014}.  Based on the presence or absence of alignment, a prediction of misassembly is made.  The {\em in silico} digestion process computationally mimics how each restriction enzyme would cleave the segment of DNA defined by the contig, returning ``mini-optical maps'' that can be aligned to the optical map for the whole genome. An important aspect of our work is that it highlights the need to use another source of information, which is independent of the sequence data but representative of the same genome, in order to identify misassembly errors. We show that optical mapping data can be used as this information source.   
 

We give results for the {\em Francisella tularensis} and loblolly pine genomes.  Each genome was assembled using various de Bruijn graph assemblers and then misassembly errors were predicted.  Our results on {\em Francisella tularensis} show that $\sequel$ correctly identifies (on average) 86\% and 80\% of locally and extensively misassembled contigs, respectively. This is a considerable improvement on existing methods, which identified (on average)  26\% and 16\% of locally and extensively misassembled contigs, respectively, in the same assemblies. The results on the loblolly pine genome assemblies show similar improvement. Lastly, our results demonstrate we are capable of significantly decreasing the false positive rate in all assemblies of {\em Francisella tularensis} and loblolly pine by incorporating optical map data into the prediction; the reduction was between 29\% and 74\%.  

\paragraph{Related Work.}  
Both amosvalidate~\cite{amos} and REAPR~\cite{reapr} are capable of identifying and correcting misassembly errors.  
REAPR is designed to use both short insert and long insert paired-end sequencing libraries, however, it can operate with only one of these types of sequencing data.  
Amosvalidate, which is included as part of the AMOS assembly package~\cite{amos2}, was developed specifically for first generation sequencing libraries~\cite{amos}. 
iMetAMOS~\cite{iMetAMOS} is an automated assembly pipeline that provides error correction and validation of the assembly.  
It packages several open-source tools and provides annotated assemblies that result from an ensemble of tools and assemblers.  
Currently, it uses REAPR for misassembly error correction. 
 
Many optical mapping tools exist and deserve mentioning, including AGORA~\cite{agora}, SOMA~\cite{soma}, and Twin~\cite{wabi2014}. AGORA~\cite{agora} uses the optical map information to constrain de Bruijn graph construction with the aim of improving the resulting assembly.   SOMA~\cite{soma} uses dynamic programming to align {\em in silico} digested contigs to an optical map.   Twin~\cite{wabi2014} is an index-based method for aligning contigs to an optical map. Due to its use of an index data structure it is capable of aligning {\em in silico} digested contigs orders of magnitude faster than competing methods.     Xavier et al.~\cite{om_mis} demonstrated misassembly errors in bacterial genomes can be detected using proprietary software.

Lastly, there are special purpose tools that have some relation to $\sequel$ in their algorithmic approach.
Numerous assembly tools use a finishing process after assembly, including Hapsembler~\cite{Donmez2011}, LOCAS~\cite{LOCAS}, Meraculous~\cite{Chapman2011}, and the ``assisted assembly'' algorithm~\cite{Gnerre2009}. Hapsembler~\cite{Donmez2011} is a haplotype-specific genome assembly toolkit that is designed for genomes that are highly-polymorphic. Both RACA~\cite{raca}, and SCARPA~\cite{scarpa} perform paired-end alignment to the contigs as an initial step, and thus, are similar to our algorithm in that respect.


%% file: methods.tex
\section{Methods}\label{sec-methods} 

$\sequel$ can be broken down into four main steps: recruitment of reads to contigs; construction of the red-black positional de Bruijn graph; misassembly error prediction; and misassembly verification using optical mapping data.   
We explain each of these steps in detail in the following subsections.

        \begin{figure}[h!]
            \centering
              	\includegraphics[width=0.6\textwidth]{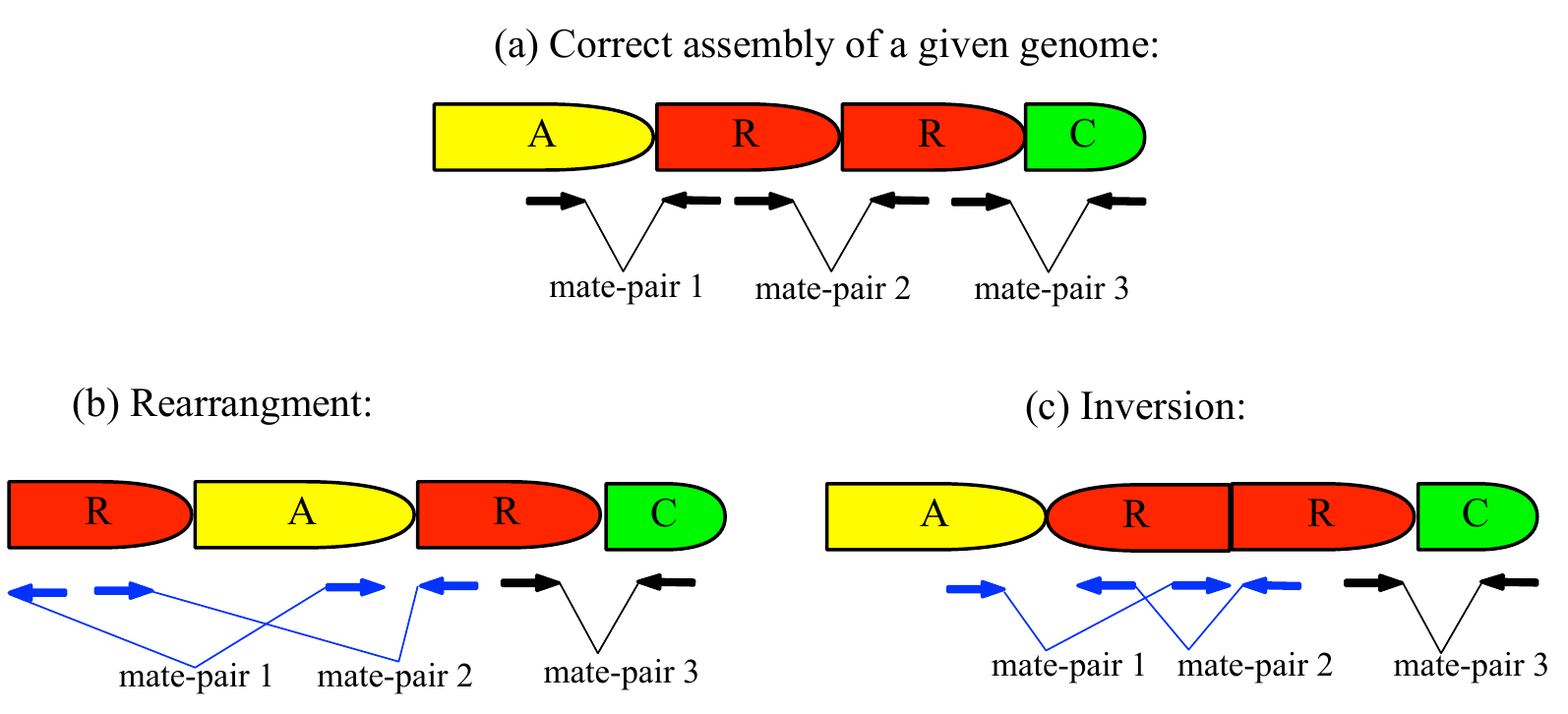}
                	\caption{An illustration of the systematic alterations that occur with rearrangements and inversions.  (a) Shows the proper read alignment where mate-pair reads have the correct orientation and distance from each other. A  rearrangement or	inversion will present itself by the orientation of the reads being incorrect, and/or the distance of the mate-pairs being significantly smaller or larger than the expected insert size. This is shown in (b) and (c), respectively.}
                	\label{fig:read_alignment_1}
        \end{figure}
       	\begin{figure}[h!]
		\centering
                	\includegraphics[width=0.6\textwidth]{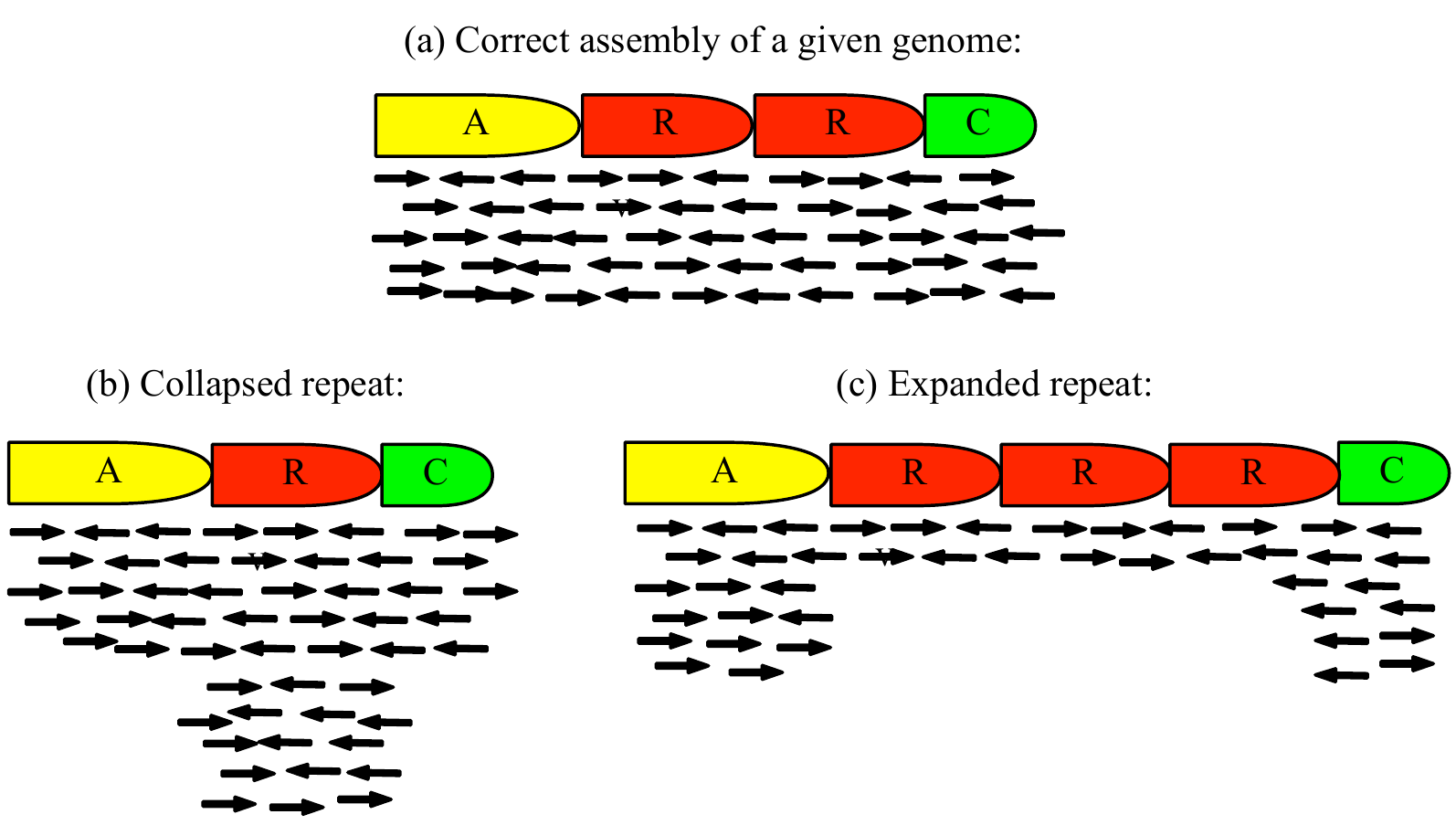}
       		\caption{An illustration of the systematic alterations that occur with collapsed or expanded repeats.  (a) Shows the proper read depth, which is uniform across the genome. A collapsed or expanded repeat will manifest as significantly lower or higher read depth; (b) shows a collapsed repeat, where the read depth being significantly greater than expected; and (c) shows a expanded repeat, where the observed read depth is significantly lower than expected.}
        		\label{fig:read_alignment_2}
        \end{figure}

\subsection{Recruitment of Reads and Threshold Calculation}  

$\sequel$ first aligns reads to contigs in order to identify regions that contain abnormal read alignments.  
Collapsed or expanded repeats will present as the read coverage being greater or lower than the expected genome coverage in the region that has been misassembled.  Similarly, inversion and rearrangement errors will present as the alignment of the mate-pairs being rearranged. Figures \ref{fig:read_alignment_1} and \ref{fig:read_alignment_2} illustrate these unconventional read alignments. More specifically, this step consists of aligning all the (paired-end) reads to all the contigs and then calculating three thresholds, $\Delta_L$, $\Delta_U$ and $\Gamma$.  The range $[\Delta_L, \Delta_U]$ defines the acceptable read depth, and $\Gamma$ defines the maximum allowable number of reads whose mate-pair aligns in an unconventional manner (e.g. inverted orientation). 
In order to calculate these thresholds, we consider all alignments of each read as opposed to just the best alignment of each read since misassembly errors frequently occur within repetitive regions where the reads will align to multiple locations.  
$\sequel$ performs this step using BWA (version 0.5.9) in paired-end mode with default parameters \cite{bwa}. 
  Subsequently, after alignment, each contig is treated as a series of consecutive 200 bp regions.  These are sampled uniformly at random $\ell$ times, and the mean ($\mu_{d}$) and the standard deviation ($\sigma_d$) of the read depth and the mean ($\mu_{i}$) and the standard deviation ($\sigma_i$) of the number of alignments where an unconventional mate-pair orientation is witnessed are calculated from these sampled regions.   $\Delta_L$ is set to the maximum of $\{0, \mu_d - 3\sigma_d\}$, $\Delta_U$ is set to $\mu_d + 3\sigma_d$, and $\Gamma$ is set to $\mu_i + 3\sigma_i$.  The default for $\ell$ is $\frac{1}{20}$th of the contig length; this can be changed via an input parameter of $\sequel$.  


\subsection{Construction of the Red-Black Positional de Bruijn Graph}

After threshold calculation, the red-black positional de Bruijn graph is constructed. For clarity, we begin by describing the {\em positional de Bruijn graph}, given by Ronen et al.~\cite{sequel}, and then define the red-black positional de Bruijn graph.  Whereas the edges in the traditional de Bruijn graph correspond to $k$-mers, the edges in the positional de Bruijn graph correspond to $k$-mers and their inferred positions on the contigs ({\em positional $k$-mers}).  Hence, the positional de Bruijn graph $G_{k, \Phi}$ is defined for a multiset of positional $k$-mers and parameter $\Phi$, and is constructed in a similar manner to the traditional de Bruijn graph using an A-Bruijn graph framework from \cite{PTT04}. Given a $k$-mer $s_k$, let $\prefix(s_k)$ be the first $k - 1$ nucleotides of $s_k$, and $\suffix(s_k)$ be the last $k - 1$ nucleotides of $s_k$.  Each positional $k$-mer $(s_k, p)$ in the input multiset corresponds to a directed edge in the graph between two positional $(k - 1)$-mers, $(\prefix(s_k), p)$ and $(\suffix(s_k), p + 1)$.  After all edges are formed, the graph undergoes a gluing operation. A pair of positional $(k - 1)$-mers, $(s_{k - 1}, p)$ and $(s_{k - 1}', p')$, are glued together into a single vertex if $s_{k - 1} = s_{k - 1}'$ and $p \in [p' - \Phi, p' + \Phi]$.  Two positional $(k - 1)$-mers are glued together if their sequences are the same and their positions are within $\Phi$ from each other. We refer to the {\em multiplicity} of a positional $(k - 1)$-mer $(s_{k - 1}, p)$ as the number of occurrences where $s_{k - 1}$ clustered at position $p$.  

$\sequel$ constructs the red-black positional de Bruijn graph from the alignment of the reads to the contigs. The red-black positional de Bruijn graph contains positional $k$-mers and is constructed in an identical way as the positional de Bruijn graph with the addition that each vertex ($(k - 1)$-mer) has an associated red or black color attributed to it that is defined using $\Delta_L$, $\Delta_U$ and $\Gamma$.  In addition to the multiplicity of each positional $(k - 1)$-mer, the number of positional $(k - 1)$-mers that originated from a read whose mate-pair did not align in the conventional direction is stored at each vertex.   When the multiplicity is less than $\Delta_L$ or greater than $\Delta_U$, or if the observed frequency of unconventional mate-pair orientation is greater than $\Gamma$, then the vertex is {\em red}; otherwise it is {\em black}.

\subsection{Misassembly Conjecture and Breakpoint Estimation}  

A red-black positional de Bruijn graph is constructed for each contig, and misassembly errors in each contig are detected by searching for consecutive red vertices in the corresponding graph.  Depth-first search is used for the graph traversal. If there are greater than 50 consecutive red vertices then the contig is conjectured to be misassembled.  The breakpoint in the contig can be determined by recovering the position of the corresponding red vertices (e.g., the positional $(k - 1)$-mers).  The number of consecutive red vertices needed to consider it misassembled can be changed via a command line parameter in $\sequel$.  Our experiments were performed with the default (e.g. 50), which corresponds to a region in the contig that has length $\geq$ 50 bp.  After this stage of the algorithm, we take contigs having regions exceeding that threshold as a set of contigs that are conjectured to be misassembled and their transitions in and out of those regions as breakpoints.

\subsection{Misassembly Verification} \label{dev}

Lastly, we use optical mapping data to verify whether a contig that is conjectured to be misassembled indeed is.  
Verification is based on the expectation that, after {\em in silico} digestion, a correctly assembled contig has a sequence of fragment sizes that is similar to that in the optical map at the corresponding locus in the genome.  In other words, an {\em in silico} digested contig should align to some region of the optical map since both are derived from the same region in the genome.
Conversely, since misassembled contigs are not faithful reconstructions of any part of the genome, when {\em in silico} digested, their sequence of fragments will likewise not have a corresponding locus in the optical map to which it aligns.  

Optical maps contain measurement error at each fragment size so some criteria is needed to decide whether variation in fragment size of an {\em in silico} digested contig and that of an optical map at a particular locus is due to variation in the size of the physical fragments or a consequence of optical measurement error.  
Due to this ambiguity, and the necessary tolerances to ensure correctly assembled contigs align to the locus in the optical map, misassembled contigs may also align to loci in the optical map, which by coincidence have a fragment sequence similar to the contig within the threshold margin of error.  
While there are various sophisticated approaches to determining statistical significance of an alignment, such as by Sakar et al.~\cite{statsigORMalign}, we use a $ \chi^2 $ model discussed by Nagarajan et al.~\cite{soma} and take the cumulative density function $\le$ 0.85 as evidence of alignment, which we found to work well empirically.

In addition, a misassembled contig only fails to align to the optical map if the enzyme recognition sequence, and thus the cleavage sites, exist in the contig in a manner that disrupts a good alignment (e.g. a misassembled contig with an inverted segment may still align if cleavage sites flank the inverted segment).
This implies that (a) some enzymes produce optical maps that have greater performance in identifying misassembly errors; and (b) alignment to the optical map is not as strong evidence for correct assembly as non-alignment to the optical map is for misassembly. 
This leads to the conclusion that an ensemble of optical maps (each made with a different enzyme) has a greater chance at revealing misassembly errors than a single optical map.  
Since acquiring three optical maps for one genome is reasonably accessible for many sequencing projects, the process of {\em in silico} digestion and alignment is repeated for three enzymes and the consensus of the alignment is taken over all of them, i.e., if two out of three times the contig did not align then it is deemed not to align (by the consensus).
A contig is deemed to be misassembled if it fails to align. The alignment is performed using Twin~\cite{wabi2014} (with default parameters) and then these results are filtered according to the $ \chi^2 $ model mentioned above.  
For our experiments, optical maps were simulated by {\em in silico} digesting reference genomes, adding normally distributed noise with a 150 bp standard deviation, and discarding fragments smaller than 700 bp.


%% file: results.tex
\section{Results} \label{sec:results} 

\subsection{Datasets} \label{subsec:data}

Our first dataset consists of approximately 6.9 million paired-end 101 bp reads from the prokaryote genome {\em Francisella tularensis}, generated by Illumina Genome Analayzer (GA) IIx platform. 
It was obtained from the NCBI Short Read Archive (accession number SRR063416). The reference genome was also downloaded from the NCBI website (Reference genome NC\_006570.2).  
The {\em  Francisella tularensis} genome is 1,892,775 bp in length. As a measure of quality assurance, we aligned the reads to the {\em Francisella tularensis} genome using BWA (version 0.5.9) \cite{bwa} with default parameters.  
We call a read {\em mapped} if BWA outputs an alignment for it and {\em unmapped} otherwise.  
Analysis of the alignments revealed that 97\% of the reads mapped to the reference genome, representing an average depth of approximately $367\times$.  

Our second dataset consists of approximately  31.3 million paired-end 100 bp reads from the loblolly pine  ({\em Pinus taeda} L.) genome~\cite{pine}.  
We downloaded the reference genome from the pine genome website~\cite{pinetreeweb} and simulated reads from the largest five hundred scaffolds from the reference using ART~\cite{art} (art illumina). 
ART was ran with parameters that simulated 100 bp paired end reads with 200 bp insert size and 50x coverage.  
The substitution error rate was reduced to one 10th of the default profile.  
The data for this experiment is available on the $\sequel$ website.  
We simulated an optical map using the reference genome for {\em Francisella tularensis} and loblolly pine since there is no publicly available one for these genomes.


\begin{table}[h!]
\begin{center}
\caption{The performance comparison between major assembly tools on the \emph{Francisella tularensis} dataset  (1,892,775 bp, 6,907,220 reads, 101 bp)  using QUAST in default mode \cite{quast}. 
All statistics are based on contigs no shorter than 500 bp. N50 is defined as the length for which the collection of all contigs of that length or longer contains at least half of the sum of the lengths of all contigs, and for which the collection of all contigs of that length or shorter also contains at least half of the sum of the lengths of all contigs.  
\# unaligned is the number of contigs that did not align to the reference genome, or only partially aligned (part).  
Total is sum of the length of all contigs. 
MA is the number of (extensively) misassembled contigs.  
local MA is the total number of contigs that had local misassemblies. 
MA (bp) is the total length of the MA contigs.  
GF is the genome fraction percentage, which is the fraction of genome bases that are covered by the assembly. 
--rr and ++rr denotes before and after repeat resolution, respectively.}
\begin{tabular}{|c|c|c|c|c|c|c|c|c|c}
\hline
\textbf{Assembler} 			&{\bf \# contigs }		& \textbf{N50}	& \textbf{Largest (bp)}	& \textbf{Total (bp) }	&\textbf{MA}	&\textbf{local MA}	& {\bf MA (bp)} 		& \textbf{GF (\%)} \\ 
 						&{\bf (\# unaligned) }		& 			& 					& 				&			& 			& 				& \\ \hline
Velvet					& 358 (3 + 35 part)		& 7,377		& 39,381				& 1,762,202 		& 11			& 36			& 84,965			& 92.09			  \\ \hline
SOAPdenovo 				& 307 (3 + 31 part)		& 8,767		& 39,989				& 2,018,158 		& 10			& 35			& 96,258			& 92.05				\\ \hline
ABySS					& 96 (1 part)  			& 27,975  		& 88,275				& 1,875,628		& 64  		& 32			& 1,330,684		& 95.87  			 	 \\ \hline
SPAdes (--rr)				& 102 (2 + 11 part) 		&  25,148		& 87,449				& 1,788,634		& 11 			& 30			& 258,309			& 92.81			 	 \\ \hline
SPAdes (+rr)				& 100 (2 + 17 part) 		&  26,876		& 87,891				& 1,797,197 		& 23 			& 31			& 497,356			& 93.75			 	 \\ \hline
IDBA						& 109 (1 + 10 part)		& 23,223 		& 87,437 				& 1,768,958		& 10			& 31			& 221,087			& 92.64  				 \\ \hline
\end{tabular}
\label{tab:ging}
\vspace{10mm}
\caption{\footnotesize{The performance comparison between major assembly tools on Loblolly pine ({\em Pinus taeda} L.) genome dataset (62,647,324 bp, 31.3 million reads, 100 bp) using QUAST in default mode \cite{quast}.}}
\label{tab:pine}
\begin{tabular}{|c|c|c|c|c|c|c|c|c|c}
\hline
\textbf{Assembler} 			&{\bf \# contigs }		& \textbf{N50}	& \textbf{Largest (bp)}	& \textbf{Total (bp) }	&\textbf{MA}	&\textbf{local MA}	& {\bf MA (bp)} 		& \textbf{GF (\%)} \\ 
 						&{\bf (\# unaligned) }		& 			& 					& 				&			& 			& 				& \\ \hline
Velvet					& 13,327 (0)			& 1,740		& 10,823				& 51,851,131 		& 0			& 0			& 0				& 62.21			  \\ \hline
SOAPdenovo 				& 16,126 (0 + 1 part)		& 7,950		& 63,004				& 57,205,817 		& 0			& 0			& 0				& 90.01				\\ \hline
ABySS					& 4,586 (16 + 89 part)  	& 37,089  		& 201,382				& 63,349,408		& 127  		& 715		& 1,391,565		& 98.17  			 	 \\ \hline
SPAdes (--rr)				& 20,671 (4 + 10 part) 	& 4,809		& 44,993				& 45,079,764		& 7 			& 11			& 65,079			& 81.30			 	 \\ \hline
SPAdes (+rr)				& 8,607 (7 + 102 part) 	& 16,957		& 108,442				& 59,730,939 		& 299 		& 57			& 3,734,609		& 94.57			 	 \\ \hline
IDBA						& 22,409 (3 + 31 part)	& 3,990 		& 40,213				& 49,765,854		& 61			& 200		& 292,769			& 79.03  				 \\ \hline
\end{tabular}
\end{center}
\end{table}

We assembled both sets of reads with a wide variety of state-of-the-art assemblers.  The versions used were those that were publicly available before or on September 1, 2014: 
SPAdes (version 3.1)~\cite{spades}; Velvet (version  1.2.10)~\cite{Zerbino:2008}; SOAPdenovo (version 2.04)~\cite{soap}; ABySS (version 1.5.2)~\cite{Simpson:2009}; and IDBA-UD (version 1.1.1)~\cite{idbaud}.
SPAdes outputs two assemblies: before repeat resolution and after repeat resolution --- we report both.
Some of the assemblers emitted both contigs and scaffolds.  We considered contigs only but note that all scaffolds had a greater number of misassembly errors. 
{\em We emphasize that our purpose here is not to compare the various assemblers, but demonstrate that all assemblers produce misassembly errors, which are in need of consideration and correction.  } 

We used Quast \cite{quast} in default mode to evaluate the assemblies.  
Quast defines misassembly error as being {\em extensive} or {\em local}.  
A (extensive) misassembled contig is defined as one that satisfies one following conditions:  (a) the left flanking sequence aligns over 1 kbp away from the right flanking sequence on the reference; (b) flanking sequences overlap on more than 1 kbp; (c) flanking sequences align to different strands or different chromosomes. 
Whereas, a local misassembled contig is one that satisfies the following conditions: (a) two or more distinct alignments cover the breakpoint; (b) the gap between left and right flanking sequences is less than 1 kbp; and the left and right flanking sequences both are on the same strand of the same chromosome of the reference genome.  
We made a minor alteration to Quast to output which contigs contain local misassembly errors.  
A contig can contain both extensive and local misassembly errors.  
Any correctly assembled contig is one that does not contain either type of error.  

\subsection{Detection of Misassembly Errors in {\em Francisella tularensis}} \label{sec:tularensis}

Table~\ref{tab:ging} gives the assembly statistics corresponding to this experiment.  
Comparable assembly results on this data were reported by Ilie et al.~\cite{sage}, though in some cases we used more recent software releases (e.g., for SPAdes).  
Note that the number of locally misassembled contigs and the number of extensively misassembled contigs is not disjoint.
A contig can be locally and extensively misassembled.   
Thus, Table \ref{tab:ging} gives the number of contigs having at least one extensive misassembly error, and the number of contigs having at least one local misassembly error.

\begin{table}[h!]
\begin{center}
\caption{The performance comparison of our method on the \emph{Francisella tularensis} dataset. 
The true positive rate (TPR) in this context is a contig that is misassembled and is predicted to be so. 
The false positive rate (FPR) is a correctly assembled contig that was predicted to be misassembled.
The TPR and FPR is given as a percentage with the raw values given in brackets}
{\setlength{\tabcolsep}{1em}
\begin{tabular}{|l|c|c|c|c|}
\hline
\textbf{Correction Method}								& \textbf{Assembler}		&{\bf MA TPR}			& {\bf local MA TPR}		& \textbf{FPR}	\\ \hline
							& Velvet				& 100\% (11 / 11)		& 100\% (36 / 36)		& 58\% (180 / 312)		\\ 
							& SOAPdenovo		& 100\% (10 / 10)		& 100\% (35 / 35)		& 63\% (165 / 263)	\\ 
 misSEQuel								& ABySS				& 100\% (64 / 64)		& 100\% (32 / 32)		& 87\% (20 / 23)			\\ 
(paired-end data only)				& SPAdes (--rr)			& 100\% (11 / 11)		& 100\% (30 / 30)		& 83\% (52 / 63)		\\ 
							& SPAdes (++rr)		& 100\% (23 / 23)		& 100\% (31 / 31)		& 86\% (49 / 57)		\\ 
							& IDBA				& 100\% (10 / 10)		& 100\% (31 / 31)		& 38\% (57 / 149) \\ \hline \hline
				
							& Velvet				& 55\% (6 / 11)			& 69\% (25 / 36)			& 24\% (76 / 312)	\\ 
							& SOAPdenovo		& 80\% (8 / 10)			& 63\% (22 / 35)			& 29\% (77 / 263)	\\ 
misSEQuel					& ABySS				& 69\% (44 / 64)		& 88\% (28 / 32)			& 13\% (3 / 23)		\\ 
(optical mapping data only)		& SPAdes (--rr)			& 91\% (10 / 11)		& 87\% (26 / 30)			& 21\% (13 / 63)		\\ 
							& SPAdes (++rr)		& 87\% (20 / 23)		& 81\% (25 / 31)			& 16\% (9 / 57)			\\ 
							& IDBA				& 90\% (9 / 10)			& 77\% (24 / 31)			& 10\% (15 / 149)		\\ 
\hline \hline
					
							& {\bf Velvet}					& {\bf 55\% (6 / 11)}			& {\bf 100\% (26 / 36)}		&	{\bf 22\% (68 / 312)}	\\ 
							& {\bf  SOAPdenovo}				& {\bf 80\% (8 / 10)}			&{\bf 84\% (21 / 35)}			&	{\bf 20\% (53 / 263)}	\\
 {\sc\bf misSEQuel}				& {\bf ABySS}					& {\bf 69\% (44 / 64)}			& {\bf 88\% (28 / 32)}			&	{\bf 13\% (3 / 23)}		\\ 
{\bf (paired-end and optical}	& {\bf  SPAdes (--rr)}				&{\bf 91\% (10 / 11)}			& {\bf 87\% (26 / 30)}			&	{\bf 19\% (12 / 63)}		\\ 
{\bf mapping data)}				& {\bf SPAdes (++rr)}				&{\bf 97\% (20 / 23)}			& {\bf 81\% (25 / 31)}			&	{\bf 16\% (9 / 57)}		\\ 
							& {\bf IDBA}					&{\bf 90\% (9 / 10)}			&{\bf  77\% (24 / 31)}			&	{\bf 9\% (14 / 149)}		\\ 
\hline \hline
							& Velvet						& 55\% (6 / 11)		& 11\% (4 / 36)				& $<$ 1\% (2 / 312)		\\ 
							& SOAPdenovo				& 20\% (2 / 10)		& 14\% (5 / 35)				& 2\% (6 / 263)	\\  
REAPR						& ABySS						& 13\% (8 / 64)		& 13\% (4 / 32)				& 4\% (1 / 23)			\\  
							& SPAdes (--rr)					& 27\% (3 / 11)		& 27\% (8 / 30)				& 5\% (3 / 63)			\\ 
							& SPAdes (++rr)				& 0\% (0 / 23)		& 19\% (6 / 31)				& 11\% (6 / 57)		\\
							& IDBA						& 40\% (4 / 10)		& 13\% (4 / 31)				& 4\% (6 / 149)			\\ 
\hline
\end{tabular}}
\label{tab:roc}
\end{center}
\end{table}

Table \ref{tab:roc} shows the results for: (a) $\sequel$ with paired-end data only; (b) $\sequel$ with optical mapping data only; and (c) $\sequel$ with both optical mapping and paired-end data in order to demonstrate the gain of combining both types of data.  
As demonstrated by these results, using short read paired-end data alone produces a high false positive rate, since it is unable to distinguish between structural variations within the genome and misassembly errors.  
This is an inherent shortcoming of short read data and demonstrates that in order to decrease the false positive rate, another source of information must be used in combination.
Optical mapping data has a much lower false positive rate and when used in combination with paired-end data, produces optimal results.  The lowest false positive rate was witnessed when both optical mapping and paired-end data were used.  In some cases, the reduction in the false positive rate was dramatic; from 87\% (ABySS, paired-end data) to 13\% (ABySS, paired-end and optical mapping data).  The true positive rate of locally misassembled contigs was between 77\% and 100\% when both paired-end and optical mapping data were used.  Lastly, true positive rate of extensively misassembled contigs was between 55\% and 100\% when both paired-end and optical mapping data were used. 

In our experiments, we iterate through combinations of three enzymes from the REBASE enzyme database \cite{roberts2010rebase} and use the set of enzymes that performed best.  
Our results demonstrate that with a good enzyme choice over half of all extensively misassembled contigs, and over 75\% of locally misassembled contigs can be identified with only a 9\%-22\% false discovery rate.
 
\subsection{Detection of Misassembly Errors in Loblolly Pine}\label{sec:pine}

The results for the loblolly pine are shown in Table \ref{tab:roc_pine}.  Both Velvet and SOAPdenovo produced zero misassembled contigs on this dataset, so we do not include them in Table~\ref{tab:roc_pine}.
$\sequel$ correctly identifies between 31\% and 100\% of extensively misassembled contigs, and between 57\% and 73\% of locally misassembled contigs.  The false positive rate was between 0.6\% and 43\%.  Although, REAPR has a lower false positive rate (between 3\% and 11\%), it is only capable of identifying a small number of extensively misassembled contigs (between 2\% and 14\%) and a small number of locally misassembled contigs (between 2\% and 27\%).  

Lastly, the restriction enzymes used in our experiments were chosen to be optimal by considering the set of all possible enzymes in the aformentioned database.  
Nonetheless, we note that if the enzyme combination was chosen at random then the expected false positive rate and true positive rate would decrease by a small fraction for majority of the assemblies considered.  
See the Appendix for prototypical ROC curves and heatmaps illustrating the density of enzyme combinations at various detection rates.

\begin{table}[h!]
\begin{center}
\caption{The performance comparison of our method on the loblolly pine dataset. 
Again, a true positive in this context is a contig that is misassembled and is predicted to be so. 
A false positive is a correctly assembled contig that was predicted to be misassembled.}
{\setlength{\tabcolsep}{1em}
\begin{tabular}{|l|c|c|c|c|}
\hline
\textbf{Correction Method}& \textbf{Assembler} 		&{\bf MA TPR}				& {\bf local MA TPR}					& \textbf{FPR}	 \\ \hline
 					& {\bf ABySS}				&  {\bf 31\% (40 / 127)}		&  {\bf 57\% (405 / 715)} 		  	 	& {\bf 43\% (1,604 / 3,754)}		 	 \\ 
{\sc\bf misSEQuel}		& {\bf SPAdes (--rr)}			&  {\bf 100\% (7 / 7)}			&  {\bf 73\% (8 / 11)}			 		& {\bf $<$1\% (135 / 20,653)	}	 	 \\ 
					& {\bf SPAdes (+rr)}			&  {\bf 67\% (199 / 299)}		& {\bf 67\% (38 / 57)}			 		& {\bf 38\% (3,117 / 8,254)} 		 	 \\ 
					& {\bf IDBA}				&  {\bf 52\% (32 / 61)}		&  {\bf 73\% (145 / 200)} 		  	 	& {\bf 19\% (4,258 / 22,150)}			 \\ 
\hline 
					& ABySS					& 7\% (9 / 127) 				& 2\% (12 / 715) 		  			& 3\% (112 / 3,754)		 \\  
 REAPR				& SPAdes (--rr)				& 14\% (1 / 7)				& 27\% (3 / 11)		 				& 6\% (1,323 / 20,653)		 	 \\ 
					& SPAdes (+rr)				& 7\% (21 / 299)			& 5\% (3 / 57)		 				& 5\% (424 / 8,254)	 \\ 
					& IDBA					& 2\% (1 / 61)				& 6\% (12 / 200)		  	 		&11\% (2,354 / 22,150)		 \\  
\hline
\end{tabular}}
\label{tab:roc_pine}
\end{center}
\end{table}

\subsection{Practical Considerations: Memory and Time} \label{mem_time}

We evaluated the memory and time requirements of $\sequel$.   Since $\sequel$ is a multi-threaded application, its wall-clock-time depends on the computing resources available to the user.  
$\sequel$ required a maximum of 8 threads, 16 GB and 1.5 hours on all assemblies of {\em  Francisella tularensis}, and a maximum of 20 GB and 2.5 hours to complete on all assemblies of loblolly pine.
Most genome assemblers require an incomparably greater amount of time and memory and thus, from a practical perspective, the requirements of $\sequel$ are not a significant increase.  
The difference in the resource requirements of $\sequel$ in comparison to modern assemblers is due to the fact it operates contig-wise rather than genome-wise and therefore, only deals with a significantly smaller portion of the data at a single time.
We conclude by mentioning that $\sequel$ is not optimized for memory and time and both could be further reduced but reimplementing the red-black positional de Bruijn graph using memory and time succinct data structures.

%% file: conclusion.tex
\section{Discussion and Conclusions} \label{sec:discussion} 

This paper describes the first non-proprietary computational method for identifying misassembly errors using short read sequence data and optical mapping data.
 Our results demonstrate: (1) a substantial number of misassembly errors can be identified in draft genomes of prokaryote and eukaryote speices; (2) our method scales to large genomes; and (3) it can be used in combination with any
 assembler and thus, making it a viable post-processing step for any assembly. 

While $\sequel$ is capable of identifying a significant percentage of misassembly errors, it does not address 
the reassembly of those the misassembled contigs. 
Correcting misassembly errors by segmenting the contigs at their breakpoints will remove the errors but will also 
reduce
the N50 
of the assembly.  
For this reason, we believe that creating a reassembly tool to correctly reassemble contigs using the misassembly information and data warrants future investigation.

While our main contributions are the computational method itself and the demonstration that optical mapping can have significant benefit for misassembly detection, optimal results are contingent upon good enzyme selection. 
Thus, we conclude by suggesting that efficient algorithmic selection of enzymes that will yield such informative optical maps in a {\em de novo} scenario is an area for interesting and important future work.



%% file: appendix.tex
\section{Appendix}

\begin{figure}[h!]
            \centering
              	\includegraphics[scale=.9]{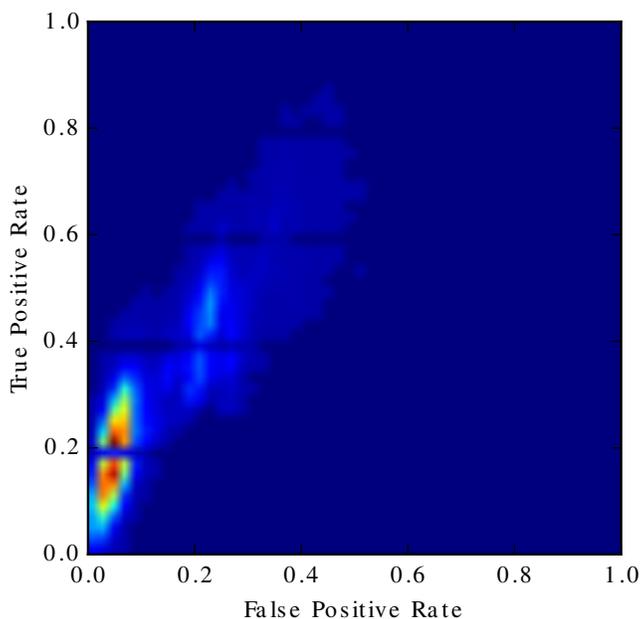}
                	\caption{ROC plot illustrating the density of optical map alignment based missassembly detection classification rates for the SOAPdenovo assembly of {\em Francisella tularensis}. The color intensity at each point indicates the number of three enzyme based classifiers having that classification rate. The plot includes results for optical maps with all three enzyme combinations using a set of 135 enzymes randomly drawn from the REBASE database.  The velvet assembly (which is not shown) has a similar pattern.  Hot spots represent the likely classification rate for enzymes choosen at random.}
                	\label{fig:soap_roc}
        \end{figure}

\begin{figure}[h!]
            \centering
              	\includegraphics[scale=.9]{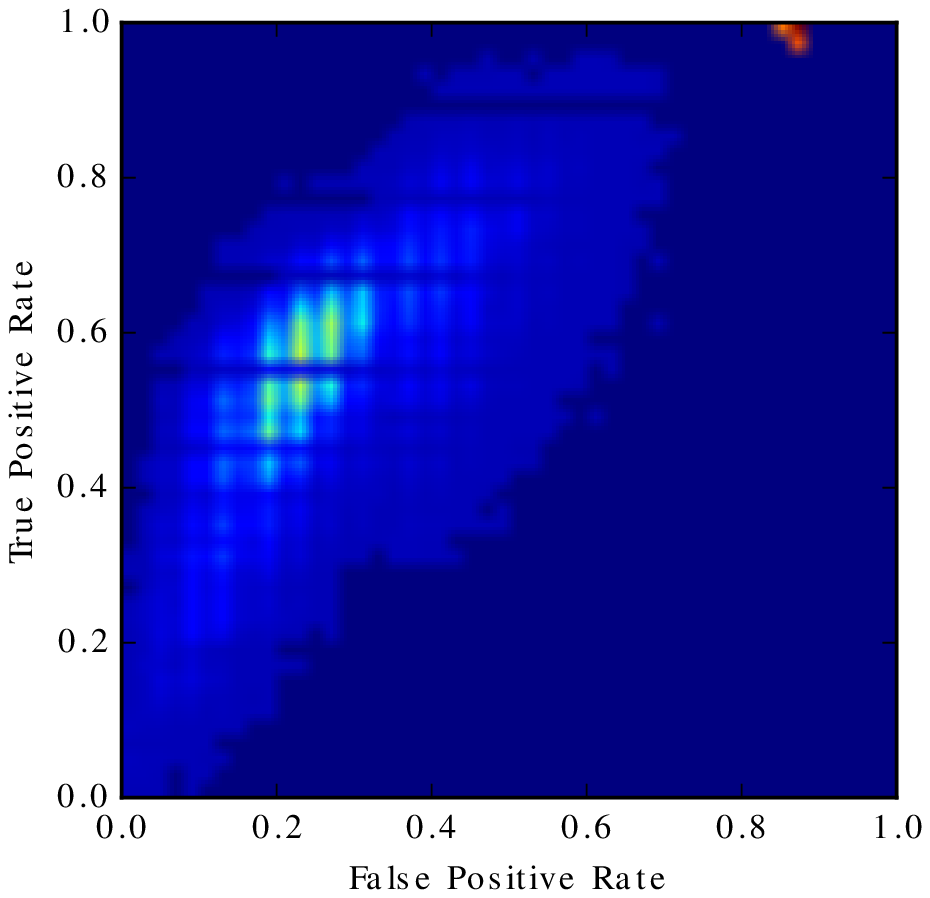}
                	\caption{ROC plot illustrating the density of optical map alignment based missassembly detection classification rates for the IDBA assembly of {\em Francisella tularensis}. The color intensity at each point indicates the number of three enzyme based classifiers having that classification rate. The plot includes results for optical maps with all three enzyme combinations using a set of 135 enzymes randomly drawn from the REBASE database.  Both SPAdes assemblies as well as ABySS (which are not shown) have a similar pattern. Hot spots represent the likely classification rate for enzymes choosen at random.}
                	\label{fig:idba_roc}
        \end{figure}

Figures \ref{fig:soap_roc} and \ref{fig:idba_roc} are the ROC plots for all (139 choose 3) enzymes for SOAPdenovo and IDBA assembly of {\em Francisella tularensis}, respectively.  The heat maps give an idea of the probability of getting a particular true positive rate and false positive rate with a specific choice of enzymes.  These plots show the sensitivity and specificity of misassembly detection using optical mapping data alone.  The paired-end sequence data was not used.  As can be seen in the plots, if a set of enzymes were chosen at random then optical mapping would still be informative and produce a meaningful classifier.  

